\documentclass[aps,pra,twocolumn,showpacs,10pt]{revtex4-1}
\usepackage{amssymb}
\usepackage{amsmath}
\usepackage{epsfig}
\usepackage{color}
\usepackage[utf8]{inputenc}
\usepackage{graphicx}
\usepackage{enumerate}
\usepackage{array}
\usepackage[normalem]{ulem}
\usepackage{soul}
\newcolumntype{M}[1]{>{\raggedright}m{#1}}

\begin{document}

\title{Surfing on protein waves: proteophoresis as a mechanism for bacterial genome partitioning}

\author{J.-C. Walter$^1$}
\author{J. Dorignac$^1$}
\author{V. Lorman$^1$}
\author{J. Rech$^2$}
\author{J.-Y. Bouet$^2$}
\author{M. Nollmann$^3$}
\author{J. Palmeri$^1$}
\author{A. Parmeggiani$^{1,4}$}
\author{F. Geniet$^1$}

\affiliation{$^1$Laboratoire Charles Coulomb (L2C), Univ. Montpellier, CNRS, Montpellier, France.}

\affiliation{$^2$ LMGM, CBI, CNRS, Univ. Toulouse, UPS, Toulouse, France.}

\affiliation{$^3$ CBS, CNRS, INSERM, Univ. Montpellier,  Montpellier, France.}

\affiliation{$^4$ DIMNP, CNRS, Univ. Montpellier, Montpellier, France.}

\date{\today}

\begin{abstract}
Efficient bacterial chromosome segregation 
typically requires the coordinated action of a three-component, fueled by adenosine triphosphate machinery called the partition complex.
We present a phenomenological model accounting for the dynamic activity of this system that is also relevant for the physics of catalytic particles in active environments. 
The model is obtained by coupling simple linear reaction-diffusion equations with a {\it proteophoresis}, or {\it ``volumetric''} chemophoresis,
 force field that arises from protein-protein interactions and provides a physically viable mechanism for complex translocation.
 This minimal description captures most known experimental observations:
 dynamic oscillations of complex components, complex separation and subsequent symmetrical positioning.
The predictions of our model are in phenomenological agreement with and provide substantial insight into recent experiments.
From a non-linear physics view point, this system explores the active separation of matter at micrometric scales with a dynamical 
instability between static positioning and travelling wave regimes triggered by the dynamical spontaneous breaking of rotational symmetry.
\end{abstract}

\maketitle
Controlled motion and positioning of colloids and macromolecular complexes
 in a fluid, as well as catalytic particles in active environments, are
 fundamental processes in physics, chemistry and biology with important implications
 for technological applications~\cite{Zottl,Marko}. 
In this paper, we focus on an active biological system for which precise experimental results are available. 
Our work is fully inspired by studies of one of the most widespread
 and ancient mechanisms of liquid phase macromolecular segregation and positioning known in nature:
 bacterial DNA segregation systems. Despite the fundamental importance of these systems in the bacterial world
 and intensive experimental studies extending over 30-years~\cite{Gerdes,Sanchez,LeGalletal-1}, 
no global picture encompasses fully the experimental observations.\\
\indent Partition systems encode only three elements that are necessary and sufficient for active partitioning:
 two proteins ParA and ParB, and a specific sequence {\it parS} encoded on DNA. 
The pool of ParB proteins is recruited as a cluster of spherical shape centered around the sequence {\it parS}, 
forming the ParB{\it S} partition complex~\cite{Sanchez}.
These ParB{\it S} cargos interact with ParA bound onto chromosomal DNA (ParA-slow)~\cite{Leonard,Bouet07}, triggering unbinding 
of ParA by inducing conformational changes through stimulation of adenosine triphosphate (ATP) hydrolysis and/or direct
 ParB-ParA contact~\cite{Vecchiarelli13}, and thereby allowing ParA diffusion in the cytoplasm (ParA-fast)~\cite{LeGalletal-1}. 
This process entails the oscillation of ParA from pole to pole and the separation of the ParB{\it S} partition complex 
into two complexes with distinct sub-cellular trajectories and long-term localization. 
Overall, these interactions result in an equidistant, stable positioning of the duplicated DNA molecules along the cell axis.\\
\indent The specific modeling of ParAB{\it S} systems falls into two categories:
 either ``filament'' (pushing/pulling the cargos, similar to eukaryotic spindle apparatus~\cite{Gerdes})
 or reaction-diffusion models~\cite{Vecchiarelli10,Vecchiarelli13,Vecchiarelli14,Ietswaart,Lim,Surovtsev,Jindal,Frey}.
Recent superresolution microscopy experiments have been unable to observe filamentous structures of ParA~\cite{LeGalletal-1,Lim},
 disfavoring polymerization-based models~\cite{Ietswaart}. 
Reaction-diffusion models have been mainly investigated numerically to describe experimental observations
 like single or multiple ParB{\it S} complex positioning.
In most cases, these models require other assumptions - such as DNA elasticity~\cite{Lim,Surovtsev} - 
as  simple  reaction-diffusion mechanisms are not sufficient to predict proper positioning.
Other reaction-diffusion models considered the dynamics of the partition complex
 on the surface of the nucleoid \cite{Vecchiarelli10,Vecchiarelli13,Vecchiarelli14,Jindal}. 
Recent experiments, however, demonstrate that partition complexes and ParA translocate through
 the interior of the nucleoid, not at its surface~\cite{LeGalletal-1}.\\
\indent Recently, in the context of the active colloids literature, there have been attempts to describe the ParAB\textit{S} system using models
 inspired by the diffusiophoresis~\cite{Anderson, AndersonBC} of active colloidal particles in solute concentration gradients~\cite{sugawara2011,Marko}. 
These works have several important limitations for applications to ParAB{\it S}, such as: rigid spherical particles (with surface reactions only),
 the steady-state approximation, only one ParA population, or reproducing equilibrium positioning only.
 The full dynamical behavior of the coupled system (ParB{\it S} cargo coupled to ParA) has thus not been elucidated.\\

 \indent Here we propose a general model of reaction-diffusion for ParA coupled to the overdamped motion of ParB{\it S}.
 Our continuum reaction-diffusion approach goes beyond the previous diffusiophoretic mechanisms~\cite{sugawara2011,Ietswaart,Jindal,Surovtsev}
 by accounting for the finite diffusion of ParA-slow and ParA-fast, as well as the interaction of ParA-slow with the entire volume of ParB\textit{S} partition complexes. 
Volumetric interactions are suggested by our recently developed ``nucleation and caging" model~\cite{Sanchez,Parmeggiani-1},
 which accounts for both the formation of ParB{\it S} and the distribution of ParB in the spatial vicinity of {\it parS} specific DNA sites~:
 the conformation of the plasmid is well described by a fluctuating polymer and the weak ParB-ParB interactions
 lead to foci of low density~\cite{Sanchez,Parmeggiani-1}. The chromosome is thus likely to enter ParB{\it S} with bound ParA-slow
 thereby allowing for volumetric interactions.
Such a volumetric interaction should also find useful applications in the field of porous catalytic particles. 
On the other hand, allowing for finite diffusion coefficients permits describing analytically the global dynamical
picture of the model, contrary to previous numerical studies often restricted to a limited range of parameters.
In particular this enables us to predict a dynamical transition between stable and unstable regimes.
 We observe that biological systems are generally close to the instability threshold.
 The ParAB{\it S} system of the F-plasmid lies just below, enabling efficient
 positioning and precursor oscillations of ParA. Other ParAB{\it S} systems (\cite{Surovtsev} and Refs. therein) could be
 just above, providing an explanation for the observed out-of-phase ParB{\it S} and ParA oscillations.
 Our model accounts for both these regimes.

\quad\\
\indent\textit{The model.} The ParA protein population is described by two coupled density fields: $u({\bf r},t)$ for the hydrolysed ParA-fast proteins,
assumed to be unbound and diffusing rapidly within the nucleoid, and $v({\bf r},t)$ for the non-hydrolysed ParA-slow molecules,
which are bound dynamically to the nucleoid and diffuse more slowly.
These two species are coupled via a system of reaction-diffusion equations:
the rapid species $u$ converts into the slow one with a constant rate $k_1$,
while the slow species $v$ is hydrolysed in the presence of the ParB\textit{S} partition complexes located on DNA,
with a rate $k_2$ (typically $k_1\approx0.02\,\rm s^{-1}$~\cite{Vecchiarelli10} and $k_2\approx68.5\,\rm s^{-1}$~\cite{Ietswaart}).
The ParB\textit{S} assemblies form 3D-foci complexes~\cite{Sanchez} and interact with the ParA-slow proteins.
The interaction probability is described by the profiles $S({\bf r}-{\bf r}_i(t))$ centered around the ParB{\it S} positions ${\bf r}_i(t)$. 
These profiles play a double role: (i) they act as catalytic sources in the reaction-diffusion equations,
triggering ParA-slow hydrolysis with the rate $k_2$ and (ii) they determine a feedback ``proteophoresis'' (volumetric) force,
 in contrast with chemophoresis forces that occur in general only at the complex surface. 
In what follows, the function $S({\bf r})$, representing an idealized density profile of ParB{\it S}, is assumed to
be symmetric with a compact support of width $\sigma$ and a unit value at its maximum.
 The dynamics of the protein population is therefore described by the coupled reaction-diffusion
equations:
\begin{align}
\frac{\partial u}{\partial t} \!=\! D_1 \, \Delta u - k_1 \, u({\bf r},t) + k_2 \, v({\bf r},t)   \sum_i S({\bf r}-{\bf r}_i(t))\,,  \nonumber \\
\frac{\partial v}{\partial t} \!=\! D_2 \, \Delta v  + k_1 \, u({\bf r},t) - k_2 \, v({\bf r},t)  \sum_i S({\bf r}-{\bf r}_i(t)) .
\label{ParA}
\end{align}
In these equations, in which we do not invoke the steady-state approximation (cf. \cite{Marko}), $D_1$ and $D_2$ represent
 the diffusion constants of the fast and slow species, respectively $u$ and $v$. The sum runs over the ParB{\it S} positions ${\bf r}_i(t)$.
The density fields are subjected to reflecting boundary conditions $\nabla u \cdot {\bf n} \rvert_{\partial V} = 0$
and $\nabla v \cdot {\bf n} \rvert_{\partial V} = 0$, where $\bf n$ is a unit vector normal to the cell boundary $\partial V$. The system described by Eqs.(\ref{ParA})
together with these boundary conditions on $u$ and $v$ ensure total ParA protein number conservation.
Note that ParA proteins can freely penetrate the partition complexes, which do not form barriers for diffusion.

The nonlinear coupling in the system is introduced by the forces driving the partition complexes,
 which are modeled as Brownian particles in an active medium.  The back reaction on each complex is described by a ``proteophoresis force''
 due to the ParA-slow concentration gradient acting on the whole volume of the complex.
In the viscous medium prevailing in a cell, we do not expect inertial terms to be important. Neglecting in the first
approximation the stochastic and confining forces, the dynamic equation for the $i^{\textrm th}$ complex then read
\begin{equation}
m \gamma \frac{d{\bf r}_i}{dt} (t) = \varepsilon \ \int_V  \nabla v({\bf r},t)
\, S({\bf r}-{\bf r}_i(t)) \, d^{3} {\bf r}.
\label{driving}
\end{equation}

Note that no direct coupling between complexes has been introduced.
The constant $\varepsilon$ represents the energy of interaction
between a single ParA-slow protein and the ParB\textit{S} partition complex. Hence, the order of magnitude of
$\varepsilon$ is a fraction of the energy released by the ATP hydrolysis ($\simeq 10 k_B T$).
The drag force coefficient $m \gamma$ is related to an effective diffusion constant of the complex $D_{\rm{pc}}$ by
the Einstein relation $m \gamma =  k_B T / D_{\rm{pc}}$. Thanks to attractive protein-protein interactions (leading to hydrolysis energy 
consumption) the interaction energy $\varepsilon$ in (\ref{driving}) is positive,
and the corresponding proteophoresis force, and resulting motion, is in the direction of increasing ParA density gradient.
 In the following, we will use the dimensionless coupling constant:
$\alpha \equiv {\varepsilon}/{m \gamma\,D_2} = ({\varepsilon}/{k_B T})(D_{\rm{pc}}/D_2)$. 
From numerical simulations it appears that the stochastic force does not affect crucially the main system dynamics.
Superresolution microscopy \cite{LeGalletal-1} indicates that the partition complex motion is
confined to the cell symmetry axis, i.e. within the bacterial nucleoid. Therefore, in the minimal model we limit the study of its dynamics to one dimension and
denote by $x$ the coordinate along the cell axis, $-L \leq x \leq L$, where $2 L$ is the cell length.\\
\indent\textit{Restoring proteophoresis force positions the partition complexes symmetrically along the nucleoid axis}.
The model provides all the necessary ingredients for proper partition complex positioning.
We first look for stationary solutions when a single partition complex is present within the cell at position $x_1$.
In order to keep the algebra simple, we approximate the profile function $S(x-x_1)$
 by a Dirac-delta distribution $\sigma\delta(x-x_1)$ \cite{footnote1}, where the amplitude $\sigma$
 is the typical interaction volume of the complex.
 The stationary solutions of Eqs.(\ref{ParA}) with reflecting boundary
conditions then reads
\begin{equation}
\left\{
\begin{array}{lcc}
u(x) = A \, \displaystyle{\frac{\cosh ( q(L+x) )}{\cosh ( q(L+x_1) )}} & \textrm{for} & -L \leq x < x_1\,,   \\
u(x) = A \, \displaystyle{\frac{\cosh ( q(L-x) )}{\cosh ( q(L-x_1) )}} &  \textrm{for} & x_1 < x \leq L\,,  \\
v(x) = C \, - \, \displaystyle{\frac{D_1}{D_2}} \, u(x)\,\,\,,  &  ~ & ~
\label{statsol}
\end{array}
\right.
\end{equation}
where $q \equiv \sqrt{k_1/D_1}$. The $x_1$ dependent constants $A$ and $C$ in (\ref{statsol}) can (see SM)
 be easily computed by the gradient discontinuity at $x_1$,
\begin{equation}
D_1 ( \partial_x u |_{x^+_1} -  \partial_x u |_{x^-_1} ) = - k_2 \, \sigma \, v(x_1),
\end{equation}
and by the conservation of the total number of ParA monomers.
 For a delta-like complex profile, the force acting on a static partition complex located at $x_1$
is proportional to the mean value of the ParA-slow density gradient at $x_1$:
\begin{multline}
F(x_1) = \frac{\varepsilon \sigma}{2}  ( \partial_x v |_{x^+_1} +  \partial_x v |_{x^-_1} )\,, \\
\!\!= \frac{1}{2}\alpha m\gamma \sigma D_1 q A\left(\tanh q (L-x_1) - \tanh q (L+x_1) \right).
\label{restoring_force}
\end{multline}

\begin{figure}[h]
\center
\includegraphics[width=.9\linewidth]{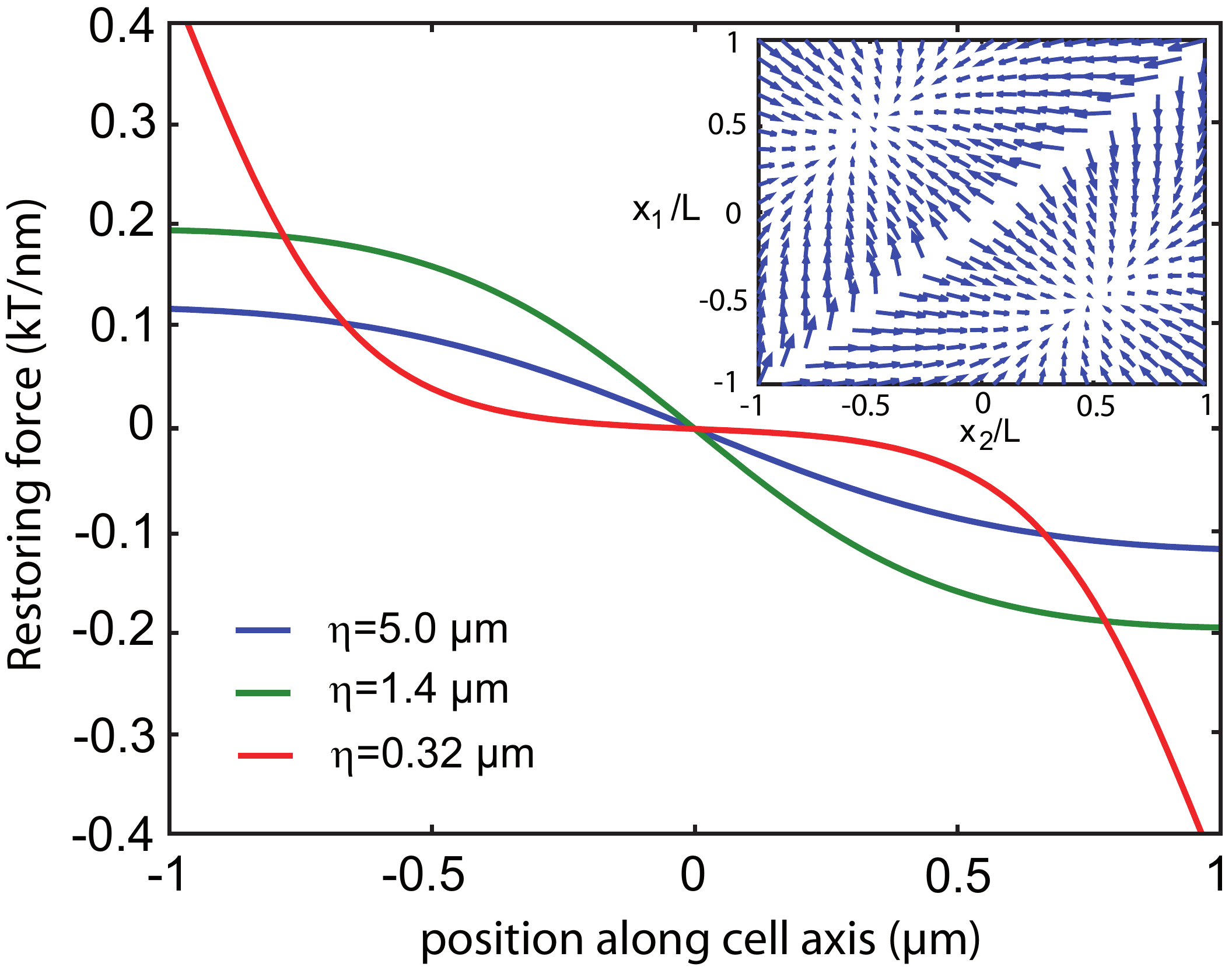}
\caption{\label{fig:Screening} (Color online) Proteophoresis force (Eq.(\ref{restoring_force})) for different values of the screening length $\eta$
 (variable $k_1$) with the other biological parameters fixed (see SM).
The curve in blue is plotted using physiological values ($k_1=0.04\,s^{-1}$) and shows a marked restoring force gradient toward mid-cell positions:
 for $\eta = 0.32$, $1.4$ and 5 $\mu$m, the  force produces a parabolic potential well of depth $\sim0$, 6 and 4~$k T$, respectively,  over a half-width  of 0.25 $\mu$m 
(note the non-monotonic behavior with the equilibrium position restoring force vanishing for both zero and infinite $k_1$, see SM). 
Inset: Proteophoresis force field in the phase space ($x_1/L, x_2/L$) of two partition complex positions.}
\end{figure}

This result shows that the unique equilibrium position of the complex is located at the cell center, i.e. $x_1 = 0$.
An important feature of the resulting force mediated by the ParA density distribution gradient is its finite range.
Its screening length, given by $\eta=1/q=\sqrt{D_1/k_1}$, is illustrated in Fig.\ref{fig:Screening},
where the force $F(x)$ is plotted for different values of $\eta$.
Clearly, the proteophoresis force, here estimated of the order of the picoNewton ($\approx0.25\,kT/$nm), is sensed by the partition complex
only if its distance to the cell boundary or to a neighboring complex is less than $\eta$.
Note that the above quasistatic (adiabatic) analysis is valid only when
the ParA distribution instantaneously adapts to the complex position (cf. \cite{Marko}). The restoring character of the force,
Eq.(\ref{restoring_force}), then makes the symmetric position $x_1=0$ stable.

\begin{figure}[ht!]
\center
\includegraphics[width=1.0\linewidth]{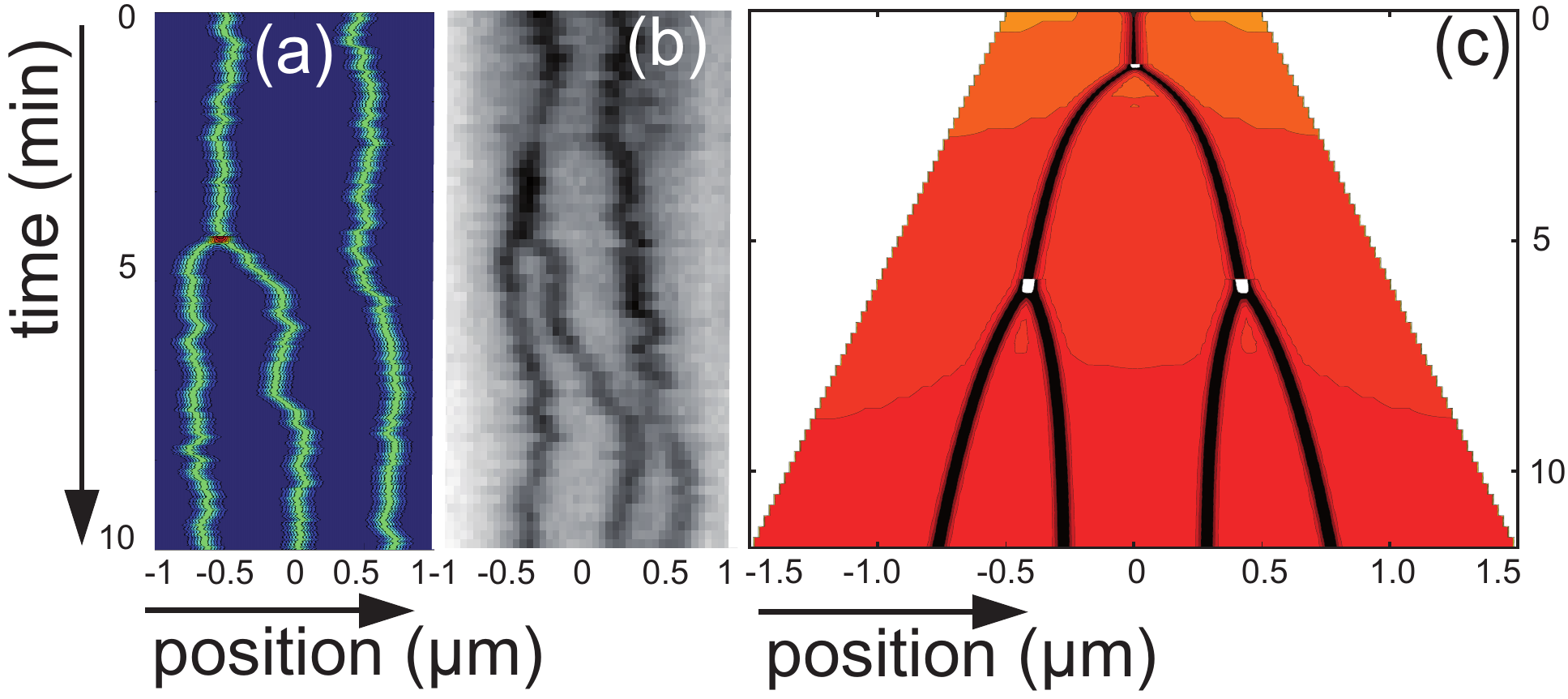}
\caption{\label{fig:kymo} (Color online)
 (a) Kymograph obtained from the model using an additional brownian force acting on ParB{\it S}: the model describes ParB{\it S} equilibrium, segregation and positioning.
 (b) Example of an experimental kymograph, obtained from 10 min. timelapse microscopy (frame every 10 sec.)
 of F-plasmids in {\it E.coli},
 displaying a segregation event from two to three ParB{\it S} over the length of the nucleoid.
 (c) Theoretical kymograph obtained with growing cell (with an average over the stochastic noise). Trajectories are similar to experimental ones \cite{LeGalletal-1}.
 For details see SM.
}
\end{figure}

For bacterial cells containing several partition complexes,
the sum over their positions in Eqs.(\ref{ParA}) generates
an effective indirect interaction among them that, together
with the boundary conditions and  protein number conservation, brings
the system to an equilibrium state with highly symmetric complex positions.
For instance, when two complexes are present within the cell (as would be the case after a DNA
replication event) the equilibrium positions are found to be located at $x_1=-L/2$ and $x_2=L/2$,
i.e. the ``1/4" and ``3/4" positions in terms of the cell axis length $2L$.
A phase portrait of the system in the $(x_1,x_2)$ coordinates (see inset of Fig.\ref{fig:Screening})
clearly indicates the stable nature of these positions. This result is in excellent agreement with
experimental observations \cite{LeGalletal-1,Glaser}, and can describe even more complex experimental situations
 with multiple ParB{\it S}, see some examples in Fig.\ref{fig:kymo}.
Interestingly, as we show below, when the evolution time scale of the ParA distribution is shorter than that of the partition complex,
 the symmetric static positions become unstable and the steady-state approximation breaks down, leading to oscillatory behavior of the complexes.

\begin{figure}
\center
\includegraphics[width=.4\textwidth]{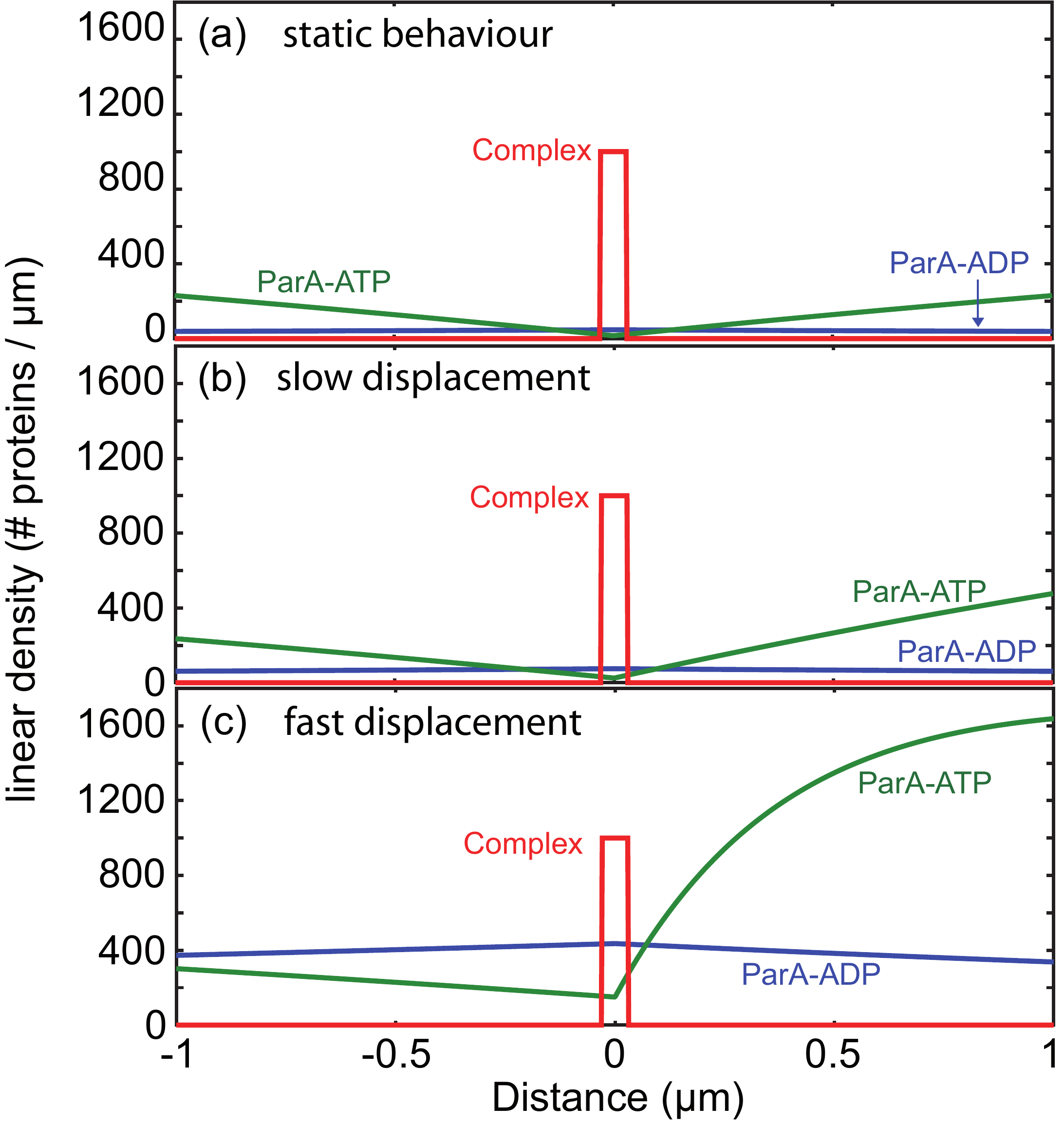}
\caption{\label{fig:TrW} (Color online) Density profile of ParA-slow $v$ (green), ParA-fast $u$ (blue) and ParB{\it S} (red).
(a) $\alpha<\alpha_c$: positioning   in the middle of the cell.
(b) weak coupling $\alpha_c \lesssim \alpha$: ParB{\it S} moves as a TW and is ''surfing" to the right on a protein wave.
(c) strong coupling $\alpha_c\ll\alpha$: large asymmetry between the two sides of ParB{\it S} implying fast ''surfing''. See SM for details.}
\end{figure}

\indent\textit{The translocation-segregation mechanism can become unstable with respect to ParA travelling waves}.
Analytical and numerical studies of Eqs. (\ref{ParA}-\ref{driving}) show that
stationary solutions (irrespective of the number of complexes) become unstable in cells where the ParA
density profiles can develop large gradients. The concentration profiles and the partition complex 
start travelling together at a constant velocity $c_{TW}$, as if partition complexes were self-propelled by
``surfing'' on the ParA distribution wave they have themselves generated (see SM) to eventually bounce back and forth
in presence of cell boundaries.
This strongly suggests the existence of travelling waves (TW) in an unbounded system
 or in finite-size cells whose length $2L$ is much larger than the screening length $\eta$.
For one complex, we look for solutions of Eqs. (\ref{ParA}-\ref{driving}) in the TW form
$u(x,t) = u(\xi); \;  v(x,t) = v(\xi)$, where $\xi=x - c_{TW}\, t$ is the wave comoving reference coordinate,
with the asymptotic conditions
$u(\xi) \rightarrow 0$  and
${v(\xi) \rightarrow v_\infty}$ when ${\xi \rightarrow \pm \infty}$.
The resulting system of ordinary differential equations admits
analytical solutions for a Dirac partition complex profile $S(x-x_1) \propto \delta(x-x_1)$.
For more general shapes, solutions are easily obtained numerically.
Typical TW-like snapshots of ParA distributions calculated for a rectangular complex profile are displayed in Fig.\ref{fig:TrW}(b,c).
The equation of motion of the partition complex (\ref{driving}) takes the form
$c_{TW}=\alpha D_2\,\int \partial_\xi v(\xi)\,S(\xi)\,d\xi$
and provides a nonlinear relation for determining the wave celerity $c_{TW}$.

The existence of travelling waves with nonzero velocity is concomitant with the loss of stability of
the equilibrium positions of the partition complexes discussed above. Thus, we distinguish two dynamical regimes:
(1) A {\it stable} regime without TWs ($c_{TW}=0$), with stable (equidistant, if more complexes are present) equilibrium complex
 positions independent of the initial conditions if the screening length $\eta$ is large with respect to the cell size, see Fig.\ref{fig:TrW}(a) and SM. 
This implies a transient translocation when the initial conditions do not correspond to stable positions. 
This regime occurs for small values both of the coupling constant $\alpha$
 (obtained, e.g.,  for large values of the limiting diffusion constant $D_2$) and the ParA concentration, $C_0$. 
When the screening length $\eta$ is small, then ParB{\it S} cargos remain at their initial positions,
 not necessarily equidistant and without interaction between complexes.
 (2) A {\it dynamical} regime ($c_{TW}\ne0$) with unstable equilibrium positions of the complexes and ParA
density oscillations in the cell corresponding to TWs in an unbounded domain, see Fig.\ref{fig:TrW}(b,c) and SM. This occurs for large values
of both $\alpha$ and the initial ParA concentration $C_0$. Since $\alpha$ is large for small values of the diffusion constant $D_2$,
 there results an apparently surprising phenomenon, namely that slower ParA-slow kinetics leads to faster complex dynamics.
This regime occurs because the ParA-slow distribution variation in time is not rapid enough to follow the partition complex and trails behind it. 
Indeed, the stability threshold corresponding to the appearence of TWs at $c_{TW}=0^+$ can be written as $V_S<V_v$, where 
$V_S$ is the escape velocity of the complex and $V_v$ the speed of spatial rearrangement of the ParA-slow distribution (see SM for details).
When $V_v > V_S$ the ParA distribution rapidly reequilibrates its symmetric profile with respect to
the complex position and the system tends to the stable stationary regime, while in the opposite case spontaneous symmetry breaking and TW behavior occur.
Using the expressions for $V_S$ and $V_v$, we obtain the stability condition in the form: $ \alpha < \alpha_c \approx 1/(\sigma C_0)$.   
This reveals that large complex sizes, interaction energies $\varepsilon$, and ParA densities,
 as well as low ParA-slow diffusion coefficients lead to the instability of the partition complex positioning.
Importantly (see SM), a biologically reasonable choice of model parameters shows that the system is
 not far below the instability threshold, leading to a not only robust but also relatively fast segregation process, in agreement with experiment.

\indent\textit{Discussion.}
 Our model for bacterial DNA segregation is able to account for the whole of the experimental phenomenology
 of segregation and positioning of the replicated DNA molecules. 
This is possible because of the careful definition of reaction-diffusion equations for the two species of ParA (slow and fast),
 coupled to the overdamped motion of the ParB{\it S} cargo. 
 
Our continuum reaction-diffusion approach significantly extends previous work
\cite{sugawara2011,Ietswaart,Jindal,Surovtsev}. Some of these \cite{Jindal,Surovtsev}
failed to observe a stable equipositioning regime because ParA-slow was not allowed to diffuse ($D_2=0$): 
thus $\alpha$ diverges, setting the system in the unstable regime.
In \cite{Surovtsev}, relative positioning occurs only with multiple cargos as a crowding
effect, whereas it is known that positioning can occur even with a single plasmid \cite{AhSeng13}, as predicted by certain modeling studies \cite{sugawara2011, Ietswaart}. 
In line with the most recent experimental findings \cite{LeGalletal-1},
we assume that partition complexes evolve within the nucleoid volume near the axis of the rod-shaped bacterial cells,
 in contrast with the translocation surface mechanism presented
in \cite{Vecchiarelli10,Vecchiarelli13,Vecchiarelli14,Jindal} performed on large surfaces coated by ParA,
lacking the confinement necessary for equipositioning. 
Our proposed mechanism integrates explicitly a volumetric interaction~\cite{Sanchez} with the partition complex (i.e. a length in 1D),
 placing the system close to the stability threshold for the biological range of parameters. 
In the case of a surface interaction, for which the volume is limited to the boundaries of the surface complex,
$\alpha_c$ would thus take much higher values.
This argument can be easily generalized to higher dimensions $D$. 
 Our approach also allows us to clarify analytically the physical mechanism at play, by going beyond the numerical
 simulations usually performed in a limited range of parameters, and to show explicitly that other effects like
 polymerization~\cite{Ietswaart} and DNA elasticity~\cite{Lim,Surovtsev} are not needed to account for segregation.\\
\indent These elements make the active system considered in our work unprecedented, with genuine size and bulk-dependent effects,
 like the emergence of a critical coupling constant controlling the stability and the TW regimes.
 Moreover, when multiple complexes are present, they generate indirect inter-complex interactions mediated solely by the ``perturbed" medium.
 This leads naturally to proper equilibrium partition complex positioning, as well as to spontaneous (left/right in 1D) symmetry breaking in the travelling wave regime. 
To our knowledge this is the first model, in the context of active bacterial segregation via ParAB{\it S} systems,
possessing very good qualitative and semi-quantitative agreement with all experimental observations,
including segregation and position control of single and multiple partition complexes (see also SM). 
 The model robustness also suggests its application to other biological processes, like macromolecule and organelle positioning in intracellular dynamics.\\
\indent Beyond its biological inspiration, this model is a novel one for active particle dynamics (accounting for ``proteophoresis'')
 and nonlinear physics with a very rich phenomenology. 
Indeed, our model falls in the class of active particles (partition complexes in the present case) which locally ``perturb" a medium
(composed here of ParA proteins) that acts back on their dynamics and thus gives rise to particle self-propulsion.
Such a behavior also provides  similarities with classical polaron systems \cite{Banyai}. 
In contrast with previous works \cite{sugawara2011,Ietswaart,Marko} on the subject, as well as on the self-propulsion of catalytic particles in active environments
under chemical gradients \cite{Zottl}, we do not  invoke specifically the well-known mechanism of diffusiophoresis (or chemiphoresis)
 \cite{Derjaguin,AndersonBC,Anderson} or  autochemotaxis, which involve surface interactions  and (possibly asymmetric) catalytic surface reactions~\cite{sugawara2011}
 coupled to surrounding hydrodynamic fluid flow relative to the particle surface (see \cite{Marko,Zottl}).
Future perspectives will include more refined comparisons with experimental observations and biological parameters and a generalization to higher dimensions.

\begin{acknowledgements}
The authors acknowledge financial support from the \textit{Agence Nationale de la Recherche} (IBM project ANR-14-CE09-0025-01)
 and from the CNRS D\'efi Inphyniti (\textit{Projet Structurant}  2015-2016). This work is also part of the program ``Investissements d’Avenir''
 ANR-10-LABX-0020 and Labex NUMEV (AAP 2013-2-005, 2015-2-055, 2016-1-024).
We thank E. Frey for informing us that he and his collaborators have used a similar approach to model the positioning of the division plane in bacteria. 
We also thank John Marko and Ned Wingreen for interesting discussions, and Martin Howard for helpful comments on the manuscript. 
\end{acknowledgements}


\begin{thebibliography}{99}

\bibitem{Zottl}  A. Zöttl and H. Stark, Emergent behavior in active colloids, J. Phys.: Condens. Matter 28, 253001 (2016).

\bibitem{Marko}
E.J. Banigan, J.F. Marko, Self-propulsion and interactions of catalytic particles in a chemically active medium, Physical Review E 93, 012611 (2016).

\bibitem{Gerdes} K. Gerdes, M. Howard and F. Szardenings. Pushing and pulling in prokaryotic DNA segregation. Cell 141, 927–942 (2010).

\bibitem{Sanchez} A. Sanchez, D.I. Cattoni, J.-C. Walter, J. Rech, A. Parmeggiani,
M. Nollmann, and J.-Y. Bouet, Stochastic self-assembly of ParB proteins builds the bacterial DNA segregation apparatus,
Cell Systems 1, 163–173 (2015).

\bibitem{LeGalletal-1} A. Le Gall, D.I. Cattoni, C. Mathieu-Demazières, L. Oudjedi, J.B. Fiche,
J. Rech, S. Abrahamsson, H. Murray, J.-Y. Bouet, M. Nollman, Bacterial partition complexes segregates within the volume of the nucleoid,
Nature Communications 7, 12107 (2016).

\bibitem{Leonard} Leonard, T.A., Butler, P.J. and Löwe, J.,
 Bacterial chromosome segregation: structure and DNA binding of the Soj dimer—a conserved biological switch.
 The EMBO journal, 24(2), pp.270-282 (2005).

\bibitem{Bouet07} Bouet J.-Y., Ah-Seng Y., Benmeradi N. and Lane D., 
 Polymerization of SopA partition ATPase: regulation by DNA binding and SopB. Molecular microbiology, 63(2), pp.468-481 (2007).

\bibitem{Vecchiarelli13}
A.G. Vecchiarelli, L.C. Hwang, K. Mizuuchi, Cell-free study of F plasmid partition provides evidence for cargo transport by a diffusion-ratchet mechanism,
Proc. Natl. Acad. Sci. USA 110, E1390--E1397 (2013).

\bibitem{Vecchiarelli10}
A.G. Vecchiarelli, Y.-W. Han, X. Tan, M. Mizuuchi, R. Ghirlando, C. Biert\"umpfel, B.E. Funnell and K. Mizuuchi,
ATP control of dynamics P1 ParA-DNA interactions: a key role for the nucleoid in plasmid partition,
Molecular Microbiology 78(1), 78-91 (2010).

\bibitem{Vecchiarelli14}
A.G. Vecchiarelli, K.C. Neuman, K. Mizuuchi, A propagating ATPase gradient drives transport of surface-confined cellular cargo,
Proc. Natl. Acad. Sci. USA 111, 4880–4885 (2014).

\bibitem{Jindal} Jindal L. and E. Emberly.
Operational principles for the dynamics of the in vitro ParA-ParB system.
 PLOS Comput. Biol. 11.12 (2015): e1004651.

\bibitem{Ietswaart}
R. Ietswaart, F. Szardenings, K. Gerdes, M. Howard,
 Competing ParA Structures Space Bacterial Plasmids
Equally over the Nucleoid, PLOS Computational Biology,  12, Vol. 10 (2014).

\bibitem{Lim} H.C. Lim, I.V. Surovtsev, B.G. Beltran, F. Huang, J. Bewersdorf, C. Jacobs-Wagner.,
 Evidence for a DNA-relay mechanism in ParABS-mediated chromosome segregation. Elife. 2014 May 23;3:e02758.

\bibitem{Surovtsev} Surovtsev IV, Campos M, Jacobs-Wagner C.,
 DNA-relay mechanism is sufficient to explain ParA-dependent intracellular transport and patterning of single and multiple cargos.
 Proceedings of the National Academy of Sciences. 2016 Nov 15;113(46):E7268-76.

\bibitem{Frey} Schumacher D., Bergeler S., Harms A., Vonck J., Huneke-Vogt S., Frey E. and S\o gaard-Andersen L.
The PomXYZ Proteins Self-Organize on the Bacterial Nucleoid to Stimulate Cell Division.
{\it Developmental Cell}, {\bf 41}, 299-314 (2017).

\bibitem{AndersonBC}
J. L. Anderson, Transport mechanisms of biological colloids., Ann N Y Acad Sci. 469, 166 (1986).

\bibitem{Anderson}
J. L. Anderson, Colloid transport by interfacial forces, Annu. Rev. Fluid Mech. 21, 61 (1989).

\bibitem{sugawara2011}
  T. Sugawara, K. Kaneko, Chemophoresis as a driving force for intracellular organization:
  Theory and application to plasmid partitioning, Biophysics, Vol. 7, pp. 77–88 (2011).


\bibitem{Parmeggiani-1} J.-C. Walter, J. Dorignac, F. Geniet, V. Lorman, J. Palmeri, A. Parmeggiani,
The caging model, a stochastic binding approach to the partition complex organization (in preparation).

\bibitem{footnote1}{It has been checked that the Dirac profile provides similar results as well as rectangular
or Gaussian profiles (see SM).}

\bibitem{Glaser} Glaser P, Sharpe ME, Raether B, Perego M, Ohlsen K, Errington J.
 Dynamic, mitotic-like behavior of a bacterial protein required for accurate chromosome partitioning.
 Genes \& Development. 1997 May 1;11(9):1160-8.

\bibitem{AhSeng13} Ah-Seng, Y., Rech, J., Lane, D., and Bouet, J. Y. (2013).
Defining the role of ATP hydrolysis in mitotic segregation of bacterial plasmids. PLoS Genet, 9(12), e1003956.

\bibitem{Banyai} L. B\'anyai, Motion of a classical polaron in a dc electric field,
Phys. Rev. Lett. 70, 1674 (1993).

\bibitem{Derjaguin}
S. Duhkin, B.V. Derjaguin, Electrokinetic Phenomena, in: E. Matijevic (Ed.), Surface and Colloid Science,
vol. 7, p. 322, Wiley–Interscience, New York, 1974.

\bibitem{RefSM} See Supplemental Material
for details on numerical simulations, microscopy experiments and biological parameters 
which includes Refs.~\cite{NumRec,Kumar,AhSeng,Bouet05,Gordon,Hyman,Diaz}.

\bibitem{NumRec} W.H. Press, B.P. Flannery, S.A. Teukolsky, T. Vetterling, Numerical Recipes in C, Cambridge University Press, 1990.

\bibitem{Kumar}
M. Kumar,M.S. Mommer and V. Sourjik
Mobility of Cytoplasmic, Membrane, and DNA-Binding Proteins in Escherichia coli
Biophysical Journal, Volume 98, February 2010, 552–559.

\bibitem{AhSeng}
Ah-Seng, Y Lopez, F., Pasta, F., Lane, D., and Bouet, J.Y (2009). Dual role of DNA in regulating ATP hydrolysis by the SopA partition protein.
 J Biol Chem 284, 30067-30075.

\bibitem{Bouet05} Bouet, J.Y., Rech, J., Egloff, S., Biek, D.P., and Lane, D. (2005)
Probing plasmid partition with centromere-based incompatibility,
{\sl Mol. Microbiol.} {\bf 55}, 511-525.

\bibitem{Gordon} Gordon, S., Rech, J., Lane, D. and Wright, A., 2004.
 Kinetics of plasmid segregation in Escherichia coli. Molecular microbiology, 51(2), pp.461-469.

\bibitem{Hyman} Hyman A.A., Weber C.A., J\"ulicher F.
 Liquid-liquid phase separation in biology,
 Annu Rev Cell Dev Biol. (2014) 30:39-58.

\bibitem{Diaz} Diaz, R., Rech, J., and Bouet, J. Y. (2015). Imaging centromere-based incompatibilities:
Insights into the mechanism of incompatibility mediated by low-copy number plasmids. Plasmid, 80, 54-62.


\end{thebibliography}
\end{document}